\documentstyle[sprocl]{article}

\input{epsf.sty}

\def\Journal#1#2#3#4{{#1} {\bf #2}, #3 (#4)}

\def\NPB{{\em Nucl. Phys.} B}

\def\PRD{{\em Phys. Rev.} D}

\def\be{\begin{equation}}
\def\ee{\end{equation}}
\def\bea{\begin{eqnarray}}
\def\eea{\end{eqnarray}}

\begin{document}

\title{HADRONIC FLUX-TUBES IN THE\\
 DUAL GINZBURG-LANDAU THEORY}

\author{\underline{Yoshiaki KOMA}, Hideo SUGANUMA and Hiroshi TOKI}

\address{RCNP, Osaka University, Ibaraki, Osaka 567, Japan
\\E-mail: koma@rcnp.osaka-u.ac.jp} 


\maketitle\abstracts{
We study the flux-tube solution in the dual Ginzburg-Landau (DGL) theory 
and apply it to the understanding of quark-gluon-plasma (QGP) formation 
in the ultra-relativistic heavy ion collision as the 
multi-flux-tube system. 
It is concluded that the annihilation process of two flux-tubes can 
contribute to the QGP formation with the energy deposition 
$\sim 4\;{\rm GeV/fm^3}$.
Furthermore, we study the glueball as the flux-tube ring by combining
the DGL theory with the string theory, and find the mass and the 
size as $\sim 1.2\;{\rm GeV}$ and $\sim 0.22\;{\rm fm}$, respectively.}

\vspace{-1.0cm}
\section{Flux-tube solution in the DGL theory}\label{sec:intro}

Dual Ginzburg-Landau (DGL) theory is an effective theory 
of non-perturbative QCD, 
and its lagrangian \cite{sst} in pure gauge is given as,
\begin{equation}
{\cal L}_{DGL}=-\frac{1}{4}(\partial_{\mu}\vec{B}_{\nu}-\partial
_{\nu}\vec{B}_{\mu})^2+\sum_{a=1}^3[|(\partial_{\mu}+ig\vec{\alpha}_a{\cdot}
\vec{B}_{\mu})\chi_a|^2\!\!-\lambda(|\chi_a|^2-v^2)^2],
\label{eqn:DGL}
\end{equation}
where $\vec{B}_{\mu}$ and $\chi_a$ denote the dual gauge field and the
monopole field, respectively, and $\vec{\alpha}_a$ is the root vector
of SU(3) algebra. 
This lagrangian has vortex-type solutions as hadronic flux-tubes 
like the Abrikosov vortex in the Ginzburg-Landau (GL) theory for 
superconductivity.
This flux-tube has constant energy per unit 
length, i.e. string tension, hence the flux-tube formation means 
the color confinement with the linear inter-quark potential. 
\par
Taking this significance of the flux-tube formation into account, 
we study the flux-tube solution
in the lagrangian (\ref{eqn:DGL}) and apply it to two topics which are
related to the non-perturbative dynamics of QCD.
One is the quark-gluon-plasma (QGP) formation scenario 
in the ultra-relativistic heavy ion collision as the multi-flux-tube system
and the other is the glueball properties, mass and size, 
as the flux-tube ring.

\section{Multi-Flux-tube System and QGP Formation}\label{sec:multi}

In the ultra-relativistic heavy ion collision, it is thought 
that the {\em multi-flux-tube system} would be realized just after 
the collision \cite{ichie}, and
whether the QGP is formed or not after the 
pre-equilibrium stage
depends on the energy deposition within the formation 
time $\sim 1\;{\rm fm/c}$.
To seek the possibility of the energy 
\begin{minipage}[hbt]{5.3cm}
deposition in such a system, 
here, we pay attention to the flux-tube interaction.
Using the DGL theory, we study the interaction energy of 
two typical two-body flux-tube systems.
One is the flux-tube and flux-tube (F-F) system and the other is 
the flux-tube and anti flux-tube (F-A) system.
These systems are fundamental systems to study the essence
of the interaction
between the flux-tubes.
We solve equations of motion so-called GL equations numerically using
finite difference method.
Here, each system is supposed to have translational invariance
along the $z$-axis.
\end{minipage}
\hspace{0.5cm}
\begin{minipage}[hbt]{5.5cm}
\vspace{-0.5cm} 
\epsfxsize=6.2cm 
\epsfysize=5.4cm
\epsfbox{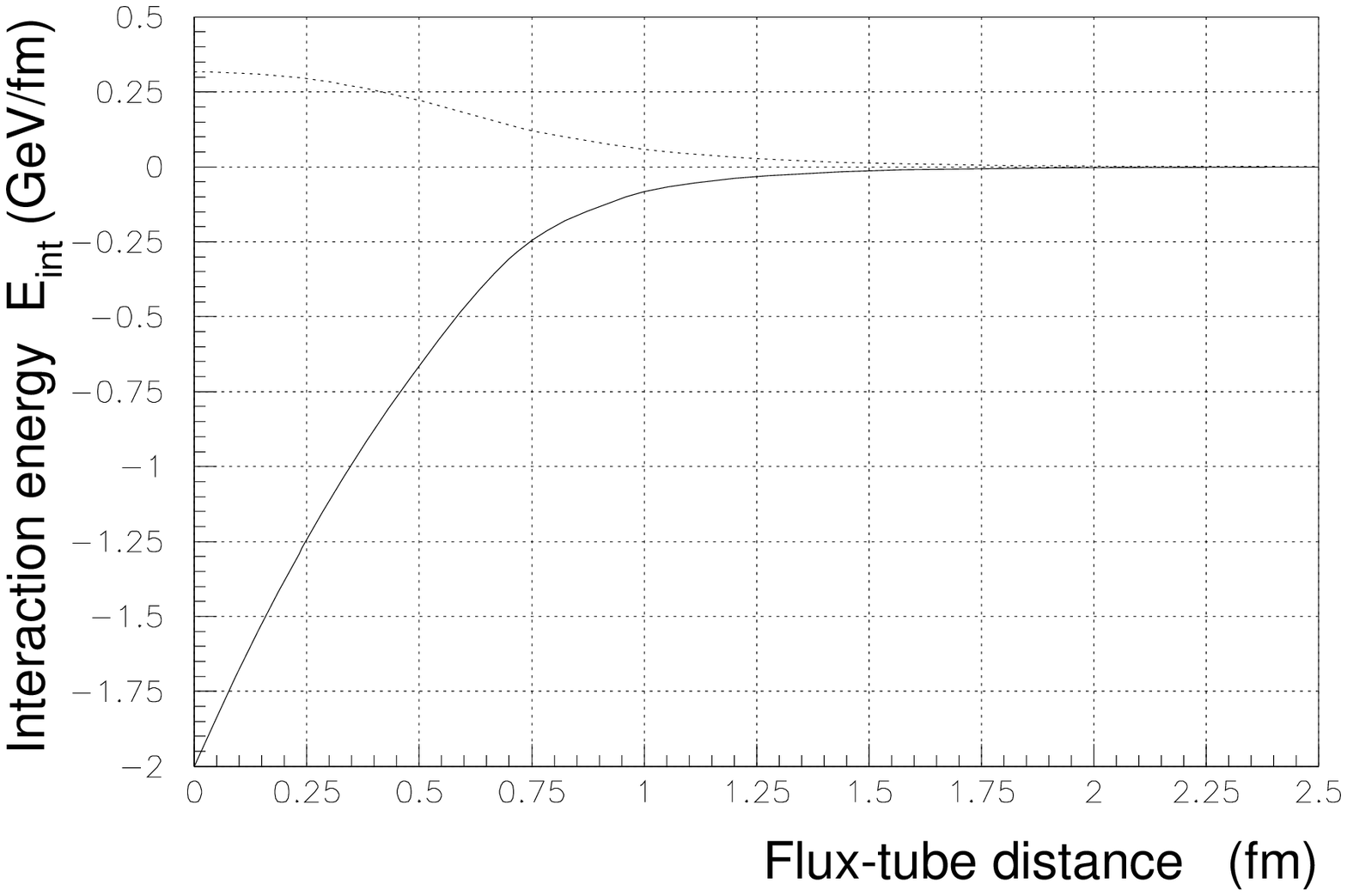}
\vspace{-0.5cm} 
{\small Fig.1: Interaction energy in the F-F system(dotted) and F-A 
system(solid) as a function of the flux-tube distance.}
\end{minipage}
\vspace{0.01mm} 
\noindent 
\par
The interaction energy of each system can be obtained 
as a function of distance between two flux-tubes shown in Fig.1, where
the interaction energy is defined as, ${\rm E}_{\rm int}\equiv 
{\rm E}_{\rm total}-2\times {\rm E}_{\rm single\;flux-tube}$. 
${\rm E}_{\rm total}$ is a total energy of the system.
These results have been calculated using the parameter set,
$v\!=\!0.10\;{\rm GeV},g\!=\!3.5\;(e\!=\!4\pi/g\!=\!3.6)$ 
and $\lambda\!=\!66$, which reproduces the string tension 
as $\sigma\!=\! 1.0\;{\rm GeV\!/fm}$.
Fig.1 shows that the interaction in the F-F system 
is repulsive, while the F-A system is strongly attractive.  
In the F-A system, flux-tube and anti flux-tube can annihilate
by its strongly attractive interaction, and 
if such a process occurs, the large energy about $4\;{\rm GeV\!/fm^3}\!
= 2\;{\rm  GeV\!/fm}\;/(\pi\!\times\!(0.4\;{\rm fm})^2)$ 
is liberated ($0.4\;{\rm fm}$ is a penetration depth of the flux-tube).
Therefore we can conclude that this annihilation process can make a
large contribution to the QGP formation.

\section{Flux-tube ring and Glueball Properties}\label{sec:ring}

Glueball is a fundamental particle of non-perturbative QCD, 
which is composed of the gluon field and considered it does not include 
a valence quark-antiquark pair.
In the DGL theory this object corresponds to the no end flux-tube, that is, 
{\em flux-tube ring}.
Therefore we can study the glueball properties by analyzing
this flux-tube ring solution in the DGL theory.
\par
We obtain the effective string tension as a function of the ring 
radius shown in Fig.2, where this value is defined as, $\tilde \sigma \equiv
{\rm E_{total}}/(2\pi R)$.
${\rm E_{total}}$ is a energy of the ring and $R$ is a ring radius,
where the system is supposed to have rotational invariance along 
the $z$-axis and the parameter set is taken just same as Sec.\ref{sec:multi}.
We find the effective string tension is reduced as the ring radius is smaller,
which suggests that the ring prefers to shrink at the classical level.
\par
However, this ring solution would be stabilized against such a collapse
by the quantum effect.
Therefore we regard the flux-tube ring as a relativistic closed string
with the effective string tension.
In the string theory \cite{string},
we can introduce the quantum effect for the string using the 
Nambu-Goto (NG) action.
By the parametrization of the ring as a circle with radius $R$,  
we can obtain the Hamiltonian from the NG action, that is, 
the energy of the closed string as, 
${\rm E}=\sqrt{P_R^2+(2\pi \tilde \sigma R)^2}$, 
%
where $P_R$ is the canonical conjugate momentum of the coordinate $R$.
Using the uncertainty relation $P_R\! \cdot \! R \!\geq\! 1$ as the 
quantum effect, 
we get the energy minimum 1.2 GeV at ring radius 0.22 fm shown in Fig.3.
We consider this energy minimum and ring radius correspond to the lower
limit of the mass and size of the glueball.
Fig.3 shows that the effective string tension lowers the glueball mass.
%
%
%
%
For more realistic calculation and to compare with the Lattice QCD 
results \cite{Tornqvist}, we must consider the motion of the ring
and extract the physical glueball state with definite quantum numbers 
`` $J^{PC}$ '' using the angular momentum projection, 
which is now under consideration.
In any case the DGL theory can provide a useful method for the study 
of the glueball. 

\begin{figure}[t]
\hspace{-0.1cm}
\begin{minipage}[hbt]{5.3cm}
\hspace{-0.1cm}
\epsfxsize=5.9cm 
\epsfysize=4cm
\epsfbox{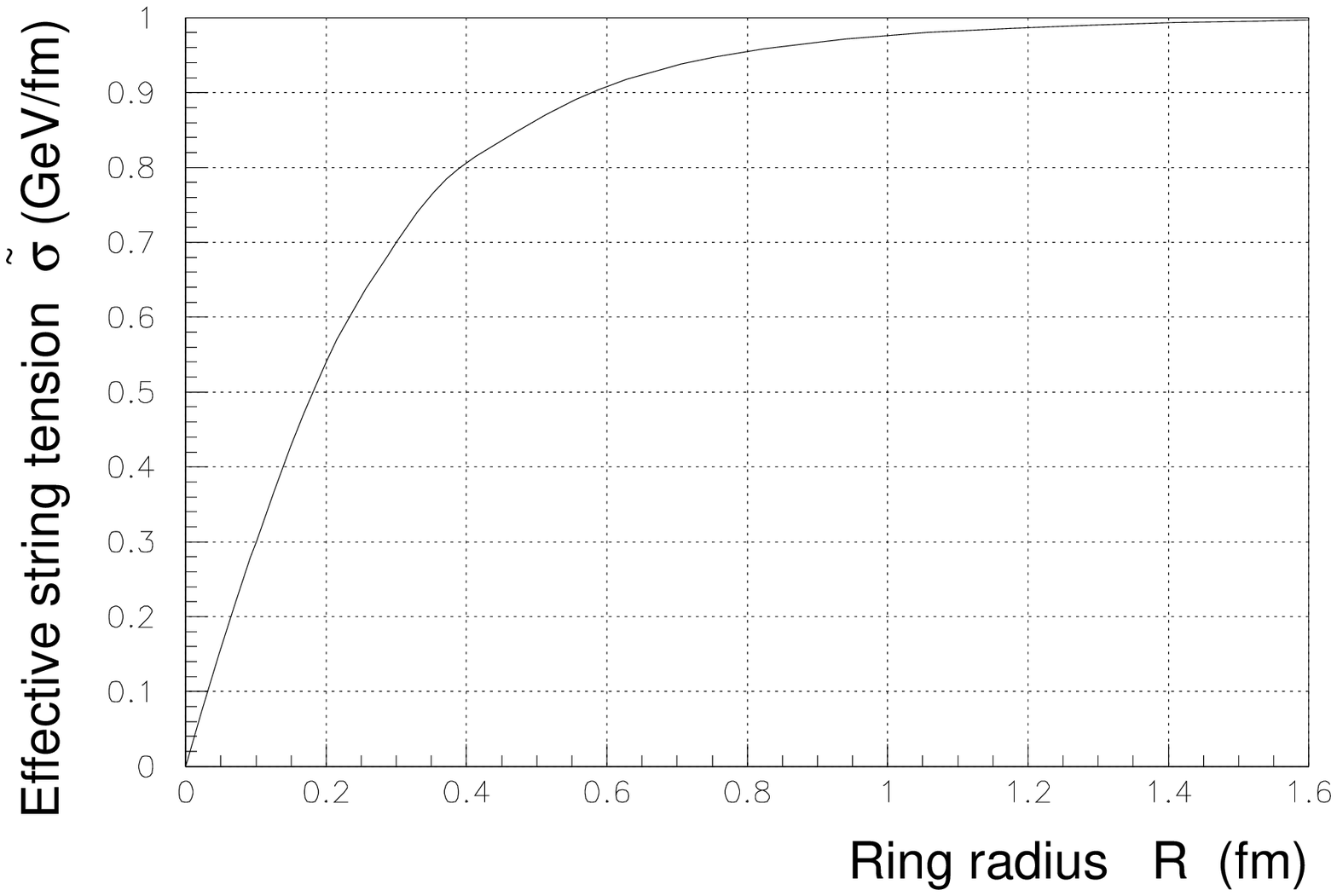}
\vspace{-0.5cm}
{\small Fig.2: Effective string tension $\tilde \sigma$
as a function of ring radius.}
\end{minipage}
\hspace{0.6cm}
\begin{minipage}[hbt]{5.3cm}
\epsfxsize=5.9cm 
\epsfysize=4cm
\epsfbox{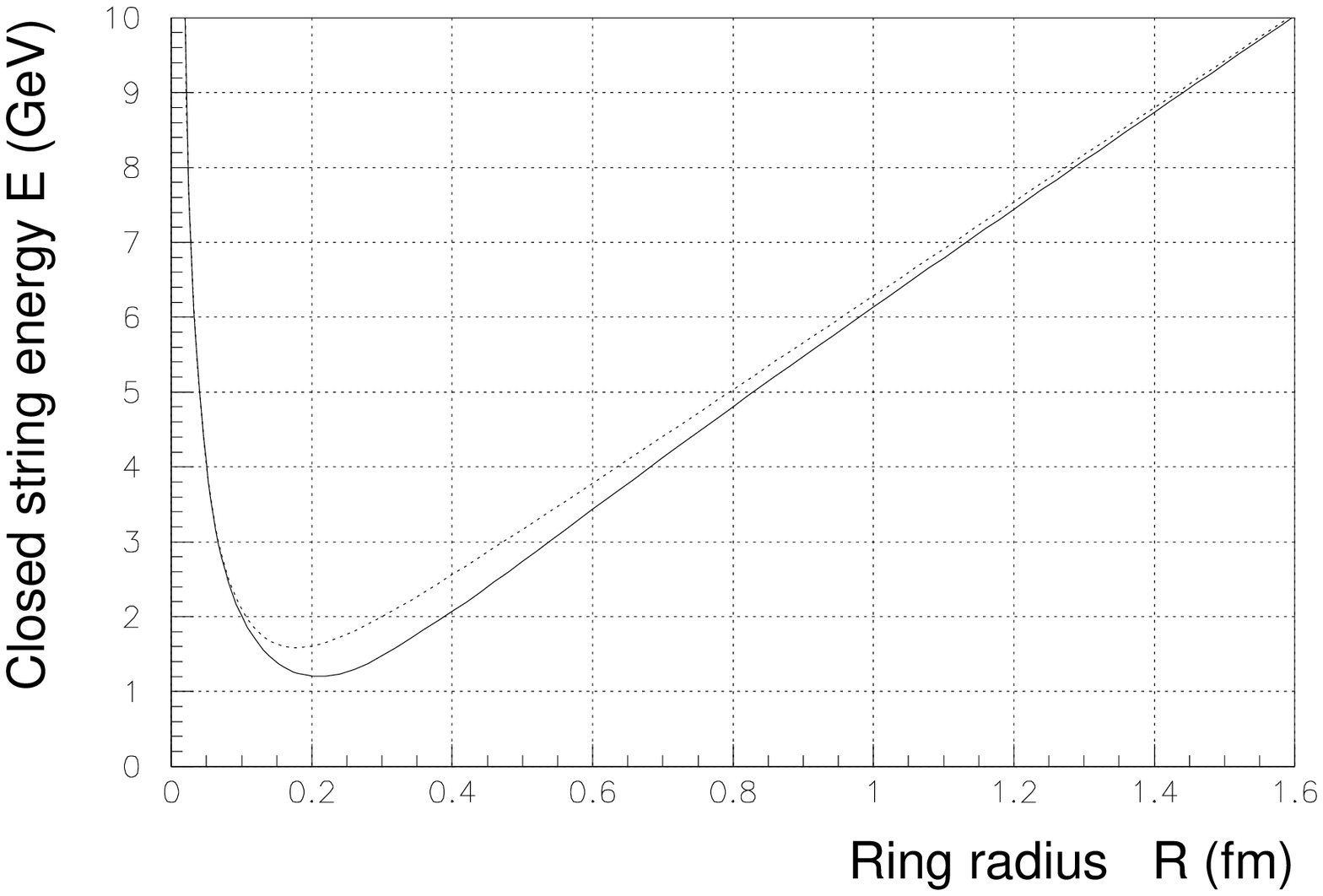}
\vspace{-0.5cm}
{\small Fig.3: Ring energy with $\tilde \sigma$(solid)
and $\sigma$(dotted) as a function of ring radius.}
\end{minipage}

\vspace{-0.2cm}
\end{figure}

\end{document}